\documentstyle[12lomcon,cite]{article}

\begin{document}

\title{THE COMMENTS ON  QED CONTRIBUTIONS 
TO  $(g-2)_{\mu}$}

\author{ A.L. Kataev \footnote{e-mail: kataev@ms2.inr.ac.ru},
\footnote{Supported in part by RFBR Grant N 05-01-00992}}

\address{Institute for Nuclear Research, 117312 Moscow, Russia}

 
\maketitle\abstracts{The comparison of  definite 
numerical values of analytically evaluated asymptotic 
expressions for order $\alpha^4$ and $\alpha^5$ QED contributions 
to the muon anomalous magnetic moment with the 
results of numerical calculations are presented. 
It is stressed that observed  agreement can be considered as the 
additional argument in favour of correctness of the recent  
direct numerical calculations.}

\section{Introduction}
The calculations  of perturbative 
QED contributions  to the  muon anomalous 
magnetic moments  $a_{\mu}$ usually  attract  
special interest.
It is   supported 
by the fact that for this low-energy quantity the interplay 
between pure QED contributions,  electroweak (EW) 
and strong interactions 
effects plays  important role. Moreover, the importance 
of taking into account strong interaction contributions 
to $a_{\mu}$ 
was    clearly demonstrated 
by comparison of the 
most recent experimental data for $a_{\mu}$  \cite{Bennett:2004pv} 
with available theoretical predictions 
(for a review see 
\cite{Passera:2004bj}). It  reveals the  
appearance of  definite, though not so clean,
deviations of theoretical and experimental results. 
The existence of this deviation  
pushed  ahead the     
desire to improve the knowledge about high order  
QED corrections  to  $a_{\mu}$.
As a result, the   numerical calculations of the sum of 
    2958 most important  five-loop diagrams, 
which depends on
$m_{\mu}/m_e$-ratio,    
were recently performed \cite{Kinoshita:2005sm}

Here  we concentrate ourselves on the comparison 
of the numerical results  for  order $\alpha^4$ and $\alpha^5$ 
contributions to  $a_{\mu}$, obtained in Refs.  
\cite{Faustov:1990zs}- \cite{Baikov:1995ui} from   
analytical high-loop contributions 
to the photon vacuum polarization function, with the existing 
results 
of direct numerical calculations.

\section{The four-loop contributions}   
The renormalization group method plays an important role in the 
high-order perturbative calculations.  
As was shown in
Ref. \cite{Lautrup:1974ic}, the coefficients of the 
${\rm ln}(m_{\mu}/m_e)$-terms in the expression for the vacuum 
polarization insertions into $a_{\mu}$ can be determined with 
the help of  the renormalization group equations in the on-shell 
scheme. The general expression for 
the contributions to $a_{\mu}$ of the  
 diagrams with  dressed by electron loops internal photon line 
in the 
muon vertex
reads 
 \begin{equation}
a_{\mu}= \frac{\alpha}{\pi}\int_0^1 dy(1-y) [d_R^{\infty}
\bigg(\frac{-y^2}{1-y}\frac{m_{\mu}^2}{m_e^2},\alpha\bigg)-1]
\end{equation}     
where $d_R^{\infty}(\alpha,x)=[1+(\alpha/\pi)\Pi(\alpha,x)]^{-1}$ 
is the asymptotic photon propagator. 
Taking into account analytical expressions 
for the constant and logarithmic contributions to the 
photon vacuum polarization function in the on-shell scheme,
one can obtain  the number 
of asymptotic  
expressions for the diagrams, contributing 
to $a_{\mu}$.  
The coefficients for 
the photon vacuum polarization function can be obtained using 
the renormalization group relations between QED $\beta$-function 
in the on-shell scheme and QED $\beta$-function in the 
momentum subtractions scheme  
which is known in the literature as the 
$\Psi$-function (see e.g. \cite{Gell-Mann:1954fq,Bogolyubov:1980nc}).
The example of application of this formalism is 
the analytical calculation of the asymptotic limit of the    
subset of  four-loop diagrams, which contain in 
the   photon line of the  external muon vertex  three-loop 
photon vacuum 
polarization contribution   with two electron loops 
({\bf Subset 1}).  
The preliminary 
consideration of Ref.\cite{Faustov:1990zs} (see 
 \cite{Tata}) 
was based on the  application of analytical  results 
for the 
four-loop contributions to the  $\Psi$-function from 
the photon vacuum polarization diagram with three electron loops 
\cite{Gorishnii:1987fy} (confirmed later on in 
\cite{Gorishnii:1990kd}) and  of  the analogous four-loop  
corrections  to the $\beta$-function in the on-shell scheme, 
taken from Ref.   
\cite{Calmet:1975rs}. These considerations 
gave    
the following 
numerical contribution 
to $a_{\mu}$   \cite{Tata} 
(see also \cite{Faustov:1990zs})
\begin{equation}
a_{\mu}(\rm{ Subset~1})= [0.923374...+ O(\frac{m_e}{m_{\mu}})]
\bigg(\frac{\alpha}{\pi}\bigg)^4
\end{equation}
which was in disagreement with the result of 
numerical calculations of Ref. \cite{Kinoshita:1990wp}, namely 
 \begin{equation}
\label{direct}
a_{\mu}(\rm{Subset~1})= 1.4416(18)
\bigg(\frac{\alpha}{\pi}\bigg)^4~~~.
\end{equation} 
As was found in Ref. \cite{Kinoshita:1990ur},   
this disagreement came from  a theoretical  error, which 
entered 
on-shell studies  of Ref. \cite{Calmet:1975rs}. 
It was    
corrected in   
Ref. \cite{Kinoshita:1990ur} and  
the new  asymptotic expression   
was obtained 
\cite{Kinoshita:1990ur,Faustov:1990zs}:
\begin{equation}
a_{\mu}(\rm{ Subset~1})= [1.452570...+ O(\frac{m_e}{m_{\mu}})]
\bigg(\frac{\alpha}{\pi}\bigg)^4~~~~~.
\end{equation}   
It differs now from Eq.(3) in the third significant digit only.
This discrepancy can be explained by the existence in Eq.(4) 
of the  $\rm{O(m_{\mu}/m_e)}$-corrections.  
The problem of comparing four-loop  asymptotic and numerical 
contributions to $a_{\mu}$  appeared again in the 
process of consideration of the subset of diagrams,
generated by the insertion into the 
internal  photon line of the muon vertex     
three-loop vacuum polarization graphs in the ``quenched'' 
approximation,    
which do not contain internal  loops.
({\bf Subset 2}).
Detailed calculational analysis of 
Ref.\cite{Broadhurst:1992za} gave    
the following numerical  
expression:
\begin{equation}
\label{BKT}
a_{\mu}(\rm{ Subset~2} )= [-0.290987...+ O(\frac{m_e}{m_{\mu}})]
\bigg(\frac{\alpha}{\pi}\bigg)^4~~~.
\end{equation}   
while the numerical results of Ref. \cite{Kinoshita:1990wp}
were 
\begin{equation}
\label{RIWIAD}
a_{\mu}(\rm{Subset~2})= -0.7945(202)
\bigg(\frac{\alpha}{\pi}\bigg)^4~~~.
\end{equation} 
The  disagreement was mostly removed in Ref. \cite{Kinoshita:1993pq}
after re-evaluating  these  diagrams
using the integration VEGAS 
routine  
(the result of Eq.(\ref{RIWIAD}) was obtained using 
the integration routine RIWIAD).
It was explained  in 
Ref. \cite{Kinoshita:1993pq} that the new 
number    
\begin{equation}
\label{VEGAS}
a_{\mu}(\rm{Subset~2})= -0.2415(19)
\bigg(\frac{\alpha}{\pi}\bigg)^4~~~
\end{equation}
differed from the one of Eq.(\ref{RIWIAD})
due to severe underestimation of errors,
which appeared in the process 
of the calculations of Ref \cite{Kinoshita:1990wp}.  
Moreover, it is closer to the asymptotic 
expression  of Eq.(\ref{BKT}), 
though definite difference of order $\rm{O(m_{e}/m_{\mu})}$ 
still remained. This difference stimulated the authors of 
Ref. \cite{Baikov:1995ui} to improve the precision of 
Eq.(\ref{BKT}), combining the asymptotic 
\cite{Gorishnii:1986pz,Gorishnii:1991hw} and 
threshold \cite{Smith:1993vp} results and applying 
the developed in Ref. \cite{Broadhurst:1993mw} variant of the Pad\'e
resummation technique. After these improvements
Eq.(\ref{BKT})  
moved in the direction 
of numerical value of  Eq.(\ref{VEGAS}). Indeed, 
the delicate calculation of Ref. \cite{Baikov:1995ui}
gave   
\begin{equation}
\label{PADE}
a_{\mu}(\rm{Subset~2})= -0.230362(5)
\bigg(\frac{\alpha}{\pi}\bigg)^4~~~.
\end{equation}  
Finally, in the work 
of Ref. \cite{Kinoshita:1998jg} the 
following new result was obtained:   
\begin{equation}
\label{new} 
a_{\mu}(\rm{Subset~2})= -0.230596(416)
\bigg(\frac{\alpha}{\pi}\bigg)^4~~~.
\end{equation} 
It is in perfect  agreement with 
 Eq.(\ref{PADE}). 
The remaining difference 
between the results of Eq.(\ref{new}) and Eq.(\ref{VEGAS}) 
was  traced to a problem  discovered in Ref.\cite{Kinoshita:1998jg}  
of rounding-off errors caused by keeping 
an insufficient number of 
effective digits  
in carrying out renormalization by numerical means. 
Thus the discovered in   Ref.\cite{Broadhurst:1992za} 
discrepancy 
between the results of analytically-oriented and 
numerical calculations 
was finally eliminated. 

\section{The five-loop contributions}
The calculations of the asymptotic contributions to $a_{\mu}$ 
from the diagrams with single  dressed internal photon line was 
continued at the five-loop level in 
Refs. \cite{Kataev:1991az}-\cite{Broadhurst:1992za}.
In particular,  Ref.\cite{Kataev:1991az} shows 
that the coefficient  of the subset  of diagrams with internal 
four-loop light-by-light scattering  graphs, 
composed from diagrams with two 
electron loops ( {\bf Subset 3}), 
is small
\begin{equation}
\label{l-b-l} 
a_{\mu}(\rm{Subset~3})= 
[-\frac{a_4^{[2,l-l]}}{2}-2.0237+ O(\frac{m_e}{m_{\mu}})]
\bigg(\frac{\alpha}{\pi}\bigg)^5~~~
\end{equation}  
where $a_4^{[2,l-l]}$ is the   constant term of 
the logarithmically 
divergent sum of the   four-loop 
light-by-light scattering contributions  
to the photon vacuum polarization 
function. 
The subset  of the  five-loop contributions to $a_{\mu}$,
formed  by insertion into  photon line 
of the sum of the four-loop ``quenched'' vacuum polarization 
graphs ({\bf Subset 4}), gives the similar   
five-loop correction \cite{Kataev:1991cp}
\begin{equation}
\label{[1]} 
a_{\mu}(\rm{Subset~4})= 
[-\frac{a_4^{[1]}}{2}-0.7334 + O(\frac{m_e}{m_{\mu}})]
\bigg(\frac{\alpha}{\pi}\bigg)^5~~~
\end{equation} 
where $a_4^{[1]}$ is the   constant term of 
the sum of the 
four-loop ``quenched''  vacuum polarization diagrams.
The contribution to $a_{\mu}$ from  the 
five-loop diagrams of {\bf  Subset 5}, which is  
generated by inserting into 
photon line of the four-loop vacuum polarization graphs 
with two electron loops, also contain still 
unknown constant terms \cite{Kataev:1991cp}:   
\begin{equation}
\label{[2]}
a_{\mu}(\rm{Subset~5})= 
[-\frac{a_4^{[2]}}{2}+0.6642 + O(\frac{m_e}{m_{\mu}})]
\bigg(\frac{\alpha}{\pi}\bigg)^5~~~.
\end{equation}
The numerical calculations of the diagrams contributing 
to  Eq.(\ref{l-b-l})-Eq.(\ref{[2]}) are 
still unavailable. 
However, there are subsets of five-loop diagrams contributing 
to $a_{\mu}$,  
which can be approximated by their  asymptotic 
expressions and were numerically calculated recently 
in the process of the work of Ref. \cite{Kinoshita:2005sm}.
These are the diagrams of six concrete subsets. 
 {\bf Subset 6} is 
formed by insertion into photon line of the muon vertex 
four-loop 
vacuum polarization graphs with three electron loops.
{\bf Subset 7} consists of   diagrams, generated  
by dressing this  photon line by  two two-loop 
vacuum polarization insertions with electron loops.
The photon line of the diagrams 
of {\bf Subset 8} contains three-loop photon vacuum 
polarization insertions with two electron loops and  
the  additional one-loop bubble.  
{\bf Subset 9} represents the sum of the diagrams,
which have ``quenched'' three-loop and one-loop subsequent 
vacuum polarization 
insertions into  photon line. 
Thus,  {\bf 
Subset 8} and  {\bf Subset 9} can be generated from 
the four-loop {\bf Subset 1} and     {\bf Subset 2} 
by inserting into their dressed  photon line the  
additional one-loop electron bubble. 
{\bf Subset 10} consists from the five-loop diagrams 
with 
the photon line,   
dressed by two-loop  photon vacuum polarization
contribution and two electron loops.   
{\bf Subset 11} is formed by inserting into photon line   
of four subsequent one-loop vacuum 
polarization contributions.
It is necessary to emphasize, that all vacuum 
polarization subgraphs of the diagrams, considered in 
this report, are  
containing {\bf electron} loops {\bf only}.
Let us   
compare the numerical analogs of    
analytical 
asymptotic five-loop contributions 
to $a_{\mu}$, calculated  in Ref. \cite{Kataev:1991cp}, 
with the concrete results \cite{Kinpriv} 
of direct numerical 
calculations  of Ref. \cite{Kinoshita:2005sm}.
The value of the  constant, related to  {\bf Subset 9}, 
was derived in the process of this work 
by more careful numerical studies  of the analytical 
result of  Ref. \cite{Kataev:1991cp}.
The asymptotic analytical expression of   
the {\bf Subset 11}
was confirmed in  
\cite{Broadhurst:1992si}. 
\newpage
{{\bf Table 1.} The  coefficients  for 
several order 
$(\alpha/\pi)^5$ 
contributions to $a_{\mu}$}.\\ 
\begin{center}
\begin{tabular}{||r|c|c||}
\hline 
Subset  of diagrams   & asymptotic result 
& numerical result  \\
\hline 
$a_{\mu}$ (\rm{Subset~6}) &  $[-\frac{\rm{a_4^{[3]}}}{2}+2.8523 
+\rm{ O(\frac{m_e}{m_{\mu}}})]$
 \cite{Kataev:1991cp}
&      2.88598(9) ~~
\cite{Kinpriv} \\
$a_{\mu}$ (\rm{Subset~7})   &  $[4.8176...  + \rm{
O(\frac{m_e}{m_{\mu}}})]$
\cite{Kataev:1991cp}   &
 4.74212(14) 
\cite{Kinpriv}  \\
$a_{\mu}$ (\rm{Subset~8)}
          &    
$[7.4491...   + 
\rm{O(\frac{m_e}{m_{\mu}}})]$  &      
 7.45270(88) 
\cite{Kinpriv}~~\\
$a_{\mu}$ (\rm{Subset~9}) & 
$[-1.3314...   +\rm{O({\frac{m_e}{m_{\mu}}}})]$ 
\cite{Broadhurst:1992za}
  & 
 -1.20841(70)
\cite{Kinpriv} \\
$a_{\mu}$ (\rm{Subset~10}) &  
$[27.7188...   + \rm{O(\frac{m_e}{m_{\mu}}})]$
\cite{Kataev:1991cp} & 
 27.69038(30) 
\cite{Kinpriv} \\
$a_{\mu}$ (\rm{Subset~11)} & 
[$20.1832...   + \rm{O(\frac{m_e}{m_{\mu}}})]$
\cite{Kataev:1991cp} & 
 20.14293(23) 
\cite{Kinpriv}\\       
\hline
\end{tabular}
\end{center}  
We  observe   satisfactory agreement 
between the  entries to the second and third column 
of Table 1.  
The existing difference     
can be explained by the  effects of the $\rm O(m_e/m_{\mu})$-
corrections.  
Indeed, for the several subsets 
the five-loop numerical 
calculations   Ref.\cite{Kinoshita:2005sm}  
are in agreement with the related analytical 
expressions obtained 
in  Ref.\cite{Laporta:1994md}  
both for the  asymptotic and $\rm O(m_e/m_{\mu})$-contributions.
Among these subsets  
are {\bf Subsets 7,10,11}. Thus, their 
asymptotic analytical results, calculated in   
Ref. \cite{Kataev:1991cp}, and confirmed in 
Ref. \cite{Laporta:1994md},
really differ from the numerical ones   
by the  leading   
$\rm{O(m_{e}/m_{\mu})}$-terms. 
In the case of the 
five-loop diagrams from {\bf Subset 9},  
the $\rm {O(m_e/m_{\mu})}$-
terms should 
also be  responsible for elimination of 
minor differences between asymptotic and numerical results.
Indeed, since   
this statement is  correct for the   
four-loop diagrams of    {\bf Subset 2}  
(compare Eq.(6) and Eq.(9) 
with Eq.(10)), it should  also be  correct  
in the case of  
five-loop diagrams 
from  {\bf Subset 9}, which have similar topological structure. 
A similar argument may also hold for  
the   
five-loop diagrams 
from  {\bf Subset 8}. 
The discussions presented 
in this summary provide additional support for     
the results 
of the  numerical  
calculations \cite{Kinoshita:2005sm} of the definite  
five-loop contributions to $a_{\mu}$. 
However, the dominant contribution of 
evaluated 
2958 diagrams  
is coming from the subset with light-by-light scattering 
electron loop internal insertion, which at present is difficult 
to check. Moreover, other  6122 diagrams 
still remained uncalculated.  
In spite of this we think  that 
that the agreement of the new  
tenth-order QED  contribution 
to $a_{\mu}$, namely  663(20)$(\alpha/\pi)^5$  
\cite{Kinoshita:2005sm},    
with the estimate,$\approx 658(\alpha/\pi)^5$ 
\cite{Kataev:2005av},  obtained by  
taking into account the results of the 10th-order 
calculations in the improved 
renormalization-group inspired  
estimates of the high-order corrections to 
is of {\bf theoretical interest} and deserve further attention.  
The current difference between phenomenological  
and experimental results  for  $a_{\mu}$ can not be described 
by the five-loop  corrections             
(for a review see \cite{Passera:2004bj}).   They  
 may  become really important for planning new  $a_{\mu}$
experiments.       

I would like to 
to thank T. Kinoshita for useful 
discussions from 1990
to the present and for the invitation to 
present the Seminar in  Cornell. 
I am grateful to S.L.  Adler for his  invitation to 
visit Institute for  Advanced Study in Princeton, 
where this summary was written.

\end{document}